# Tunable Sorting of Mesoscopic Chiral Structures by External Noise in Achiral Periodic Potentials

**Jie Su, Hui-Jun Jiang[*] and Zhong-Huai Hou***

*Department of Chemical Physics & Hefei National Laboratory for Physical Sciences at the Microscale, University of Science and Technology of China, Hefei 230026, China.*

Correspondence and requests for materials should be addressed to the author: H.-J.J. (email: hjjiang3@ustc.edu.cn) or Z.-Z.H. (email: hzhlj@ustc.edu.cn)

**Abstract:**

Efficient chirality sorting is now highly demanded to separate assembled mesoscopic chiral structures which are of very special physical properties rather than their achiral counterparts or those at the single-particle level. However, the efficiency of conventional methods usually suffers from the thermal or external noise. Here, we propose a mechanism utilizing external noise to attain a tunable sorting of mesoscopic chiral particles in an achiral periodic potential. The complete chirality-separation stems from the path selection by a noise-induced biased flux in a nonequilibrium landscape. Such mechanism provides a practicable way to control the motion of chiral particles by simply adjusting the noise intensity, which is demonstrated by simultaneous separation of several kinds of enantiomorphs with different degrees of chirality. The robustness and generalizability of noise-tuned chirality sorting is further verified in systems with other types of periodic potentials or spatially/temporally correlated noise.

**Introduction**

Chirality is a property of mirror asymmetry essential in several branches of science[1-3]. Very recently, explosive attention has been paid to mesoscopic chiral structures assembled from micro- or meso-blocks[4-10]. Such assembled chiral structures are shown to be of very special optical, electric or magnetic properties rather than their achiral counterparts or those at the single-particle level[4-8]. Since the assembled products are usually mixtures of enantiomorphs[4,8], methods to efficiently sort such chiral structures then receive great research interest.

So far, several types of chirality-sorting methods have been proposed. Usually, a system with intrinsic chirality along the separation direction, for instance, asymmetric shear flow[11-16], helical flow field[17-19] and some materials such as chirality-separation sieves[20-22], can be utilized to sort chiral objects. Interestingly, periodic potentials without intrinsic chirality can also be used to separate enantiomorphs, thus providing a more convenient way for chirality sorting. In the pioneer work done by de Gennes[23], a macroscopic chiral crystal was found to glide in a direction differing slightly from the axis of maximum slope when it was slipping over an inclined solid support. For separation of smaller chiral objects such as macromolecules or assembled mesocopic structures where thermal fluctuation is nonnegligible, it was argued that the fluctuation would destroy this effect[23]. In terms of this, many efforts have been paid to search for efficient chirality sorting methods against thermal noise[24-26]. Speer *et al.* demonstrated that the two chiral counterparts can even move into opposite directions with remarkable persistence against thermal noise with the help of periodic potentials[24]. Similarly, particles that only differ by their chirality were also found to migrate along different directions when driven by a steady fluid flow through a square lattice of cylindrical posts when thermal noise presents[25]. Furthermore, it has also been reported that periodic potentials can display not only chirality separation, but also the ability to steer particles to arbitrary locations[26]. One should note that, for a real system, thermal fluctuations are inevitable, which are related to the dissipation process such as friction via fluctuation-dissipation theorem. Such thermal fluctuations may be regarded as "internal" to the system dynamics. In addition, the system's dynamics may also be influenced by noises from the environment or external fields. However, many studies have shown that external noise can play constructive roles in nonlinear dynamic systems and lead to counterintuitive phenomena, such as stochastic resonance[27-31]. Therefore, it is very interesting to ask whether external noise can also be favorable for chirality sorting rather than destroying it. Since external noise is unavoidable in real systems and may be tuned systematically, the answer to this question may provide new methods to achieve tunable chirality sorting.

Here, we establish a two dimensional noise-tuned system consisting of assembled chiral particles driven by flowing fluid in an achiral periodic potential. As a result, an optimal chirality sorting with complete chirality-separation and an interesting rollover phenomenon are observed by solely adjusting the noise intensity. Analysis based on nonequilibrium landscape and flux theory reveals that, within the nonequilibrium landscape, a noise-induced biased flux would guide particles of a given chirality to move along a selected direction, leading to the optimal chirality separation with 100% selectivity as well as the interesting rollover of chirality sorting. Detailed analysis on dynamical trajectories finds that, the biased flux is generated by a noise-induced path transition, which may be further associated with the intrinsic dynamical asymmetry of enantiomorphs. More interestingly, the selected direction shows a quantitative dependence on the noise intensity, revealing a tunable motion of chiral particles by external noise. Based on the mechanism, simultaneous separation of several kinds of enantiomorphs with different degrees of chirality is successfully realized. The robustness and generalizability of noise-tuned chirality sorting in systems with other types of periodic potentials or spatially/temporally correlated noise are further verified. Thus, our method provides a conceptually new and practicable way for tunable chirality-sorting in real systems.

**Model and Methods**

We consider a mesoscopic 2-dimensional chiral structure as an equilateral trilateral particle with side length $l_0$ assembled by three rigidly coupled nanoparticles located at $\boldsymbol{r}_i$, i=1,2,3 (Figure 1). The chirality of the particle is realized by setting the three nanoparticles to be of different sizes, which consequently results in different friction coefficients $\gamma_i$[12,17]. Taking $\gamma_1 < \gamma_2 < \gamma_3$, we consider a particle is of (+) chirality if the 1st, 2nd and 3rd nanoparticles (nodes) are arranged counterclockwise, and is of (-) chirality for clockwise arrangement (the inset in Figure 1). The state of such a rigid particle can then be described by the position of its friction center $\boldsymbol{R} = \sum_{i=1}^{3} \gamma_i \, \boldsymbol{r}_i / \sum_{i=1}^{3} \gamma_i$, and by the orientation angle $\phi$ between the vector $\boldsymbol{r}_1 - \boldsymbol{R}$ and the X axis. The position of each nanoparticle can be determined by $\boldsymbol{r}_i = \boldsymbol{R} + \boldsymbol{q}_i(\phi) = \boldsymbol{R} + \boldsymbol{O}(\phi)\boldsymbol{q}_i^{(0)}$, where $\boldsymbol{q}_i$ is the vector pointing from $\boldsymbol{R}$ to node-i, $\boldsymbol{q}_i^{(0)}$ denotes its value in a reference configuration with $\phi = 0$, and $\boldsymbol{O}(\phi)$ is a rotation matrix whose elements are $O_{11} = O_{22} = \cos(\phi)$ and $O_{21} = -O_{12} = \sin(\phi)$. Then the friction center evolution can be calculated by the motions of the three nanoparticles[17,24], i.e., the particle obeys the coupled Langevin equations as follows,

$$\frac{d\boldsymbol{R}(t)}{dt} = \frac{\sum_{i=1}^{3} \boldsymbol{F}(\boldsymbol{r}_i) + \sum_{i=1}^{3} \gamma_i \boldsymbol{v}^{fl}(\boldsymbol{r}_i)}{\sum_{i=1}^{3} \gamma_i} + \boldsymbol{\xi}(t) + \boldsymbol{\zeta}(t), \tag{1}$$

$$\frac{d\phi(t)}{dt} = \frac{\boldsymbol{e}_z \cdot \sum_{i=1}^{3} \boldsymbol{q}_i(\phi) \times [\boldsymbol{F}(\boldsymbol{r}_i) + \gamma_i \boldsymbol{v}^{fl}(\boldsymbol{r}_i)]}{\sum_{i=1}^{3} \gamma_i |\boldsymbol{q}_i|^2} + \xi_\phi(t) + \zeta_\phi(t). \tag{2}$$

Herein, $\boldsymbol{F}(\boldsymbol{r}_i) = -\nabla U(\boldsymbol{r}_i)$ denotes the force exerted to node-i, where

$$U(\boldsymbol{r}) = C\gamma \sum_{j=1}^{3} c_j \cos(K\hat{k}_j \cdot \boldsymbol{r} + \delta_j) \tag{3}$$

is the periodic potential taken as a superposition of three standing waves[26], with $C$ the potential strength, $K$ the inverse spatial scale, and $\hat{k}_j$, $c_j$, $\delta_j$ respectively the unit vector, ratio and phase offset associated with the j-th standing wave ($\sum_{j=1}^{3} c_j^2 = 1$). Such a potential may be realized by an external electric or optical field[26,32]. We fix $c_1 = 0.256$, $c_2 = c_3 = 0.683$, $\hat{k}_1 = (1,0)$, $\hat{k}_2 = (-0.5, \sqrt{3}/2)$, $\hat{k}_3 = (0.5, \sqrt{3}/2)$, $\delta_1 = 0$, $\delta_2 = 2.49$, and $\delta_3 = 3.79$, such that the maxima and minima of the potential lie on interpenetrating triangular lattices (Figure 1). The potential has mirror symmetry parallel to the X axis, but none parallel to the Y axis.

In Eq. (1), $\boldsymbol{v}^{fl}(\boldsymbol{r}_i) = v_0 \boldsymbol{e}_x$ denotes a velocity field of a fluid with constant speed $v_0$, which provides driven force for particle movement along the X axis. $\boldsymbol{\xi}(t)$ is thermal noise which can be expressed as an independent Gaussian white noise satisfying the fluctuation-dissipation relation $\langle \xi_\mu(t)\xi_\nu(t') \rangle =$ $2\gamma_c^{-1}k_BT\delta(t-t')\delta_{\mu\nu}$, where the subscript $\mu(\nu)$ denotes the component along the X(Y) axis, $k_B$ is the Boltzmann constant, $T$ denotes the temperature, and $\gamma_c = \sum_{i=1}^{3}\gamma_i$ represents the total friction coefficient. $\boldsymbol{\zeta}(t)$ is an external noise with $\langle\zeta_\mu(t)\zeta_\nu(t')\rangle = 2D_e\delta(t-t')\delta_{\mu\nu}$, where the intensity $D_e$ can be tuned, externally. Eq. (2) represents the rotation of the chiral particle, where torques are exerted by $\boldsymbol{F}$ and $\gamma_i\boldsymbol{v}^{fl}$. The scalar $|\boldsymbol{q}_i|$ is the norm of $\boldsymbol{q}_i$, as well as $\xi_\phi(t)$ and $\zeta_\phi(t)$ are the internal and external rotational fluctuation respectively satisfying $\langle\xi_\phi(t)\xi_\phi(t')\rangle = 2k_BTr_\gamma^{-2}\delta(t-t')$, $\langle\zeta_\phi(t)\zeta_\phi(t')\rangle = 2D_e\gamma_c r_\gamma^{-2}\delta(t-t')$, where $r_\gamma = (\sum_{i=1}^{3}\gamma_i|\boldsymbol{q}_i|^2)^{1/2}$ is independent of $\phi$ and represents an invariant property of the molecule[17].

In simulations, parameters are made dimensionless by using $l_0$, $\gamma_1$ and $v_0$ as the basic units. Accordingly, the basic unit for time is $l_0/v_0$ that for energy is $l_0v_0\gamma_1$. We fix $K = 0.3$, $C = 6.67$ and $k_BT = 10^{-5}$ all through the present work. For consistency, all of the following results are obtained from particles running for a long time $t = 10^6$ with $2\times10^4$ randomly chosen initial states with different orientations near the origin. In the following, we mainly focus on the dynamics of (+) particles, since the behaviors of (-) particles are the same except that their motions are symmetric to those of (+) ones with respect to the X-axis.

**Results and Discussion**

We now investigate how the external noise influence chirality sorting of particles with $\gamma_1 = 1.0$, $\gamma_2 = 1.5$ and $\gamma_3 = 2.0$. Typical trajectories for $D_e = 0$, $10^{-4}$ and $0.1$ are shown in the insets of Figure 2, respectively. Without noise ($D_e = 0$, the top-left inset in Figure 2), both enantiomorphs can move along two different paths, one of which along positive Y-direction (named as the positive path, movie S1), and the other with a negative one (the negative path, movie S2). Nevertheless, trajectories for (+) and (-) particles are mixed with each other, so that the two enantiomorphs cannot be separated. Quite interestingly, for a moderate noise intensity $D_e = 10^{-4}$(the bottom inset in Figure 2), all of the (+) particles move along the positive path while all (-) particles along the negative one, resulting in two well separated clusters of particles, wherein the two enantiomorphs are nearly 100% sorted. If the noise intensity becomes large enough, for instance $D_e = 0.1$ (the top-right inset in Figure 2), (+) and (-) particles are mixed again.

The above observation clearly indicates that external noise can induce chirality sorting, and there exists an optimal level of noise intensity that may most favorably enhance the sorting selectivity. In order to quantitatively characterize such effect, we may introduce an order parameter $S$ to measure the chirality selectivity as

$$S = \int_0^\infty P_{(+)}(Y)dY - \int_0^\infty P_{(-)}(Y)dY. \tag{4}$$

where $P_{(\pm)}(Y)$ denotes the probability distribution of the final Y-position for the ($\pm$) particles. For $S > 0(S < 0)$, (+) particles distribute more(less) above X axis than (-) particles. $S = 0$ means that no chirality sorting can be observed, while for $S = \pm 1$ the two enantiomorphs can be separated completely. The obtained $S$ as a function of $D_e$ is presented in Figure 2. It is observed that chirality selectivity depends non-monotonically on the intensity of the external noise, and at an optimal noise intensity, a complete sorting with 100% chirality selectivity can be realized.

In order to understand the underlying mechanism of the interesting observations aforementioned, we now try to figure out the nonequilibrium potential landscape and flux for the periodic positions of chiral particles in a lattice of the potential at different noise intensities. Generally, the evolution equation of the friction center can be written as $d\overline{\mathbf{R}}/dt = f(\overline{\mathbf{R}}) + \xi_{\overline{\mathbf{R}}}(t)$, where $\overline{\mathbf{R}} = (\overline{X}, \overline{Y})$ is the periodic position of the friction center in a lattice of the potential, i.e., $\overline{X}(t) = mod(X(t), L_X)$, $\overline{Y}(t) = mod(Y(t), L_Y)$ with $L_X$ and $L_Y$ the periodic length of the given potential. $f(\overline{\mathbf{R}})$ is a deterministic force and $\xi_{\overline{\mathbf{R}}}(t)$ is a random force satisfying $\langle \xi_{\overline{\mathbf{R}}}(t)\xi_{\overline{\mathbf{R}}}(t')\rangle = 2D\delta(t - t')$ with $D$ the corresponding diffusion coefficient. Via the Fokker-Plank equation for the probability distribution $P(\overline{X}, \overline{Y})$, the nonequilibrim potential landscape and flux can then be well defined as $U_{neq}(\overline{X}, \overline{Y}) = -\ln P_{SS}$, $J_{SS,\overline{X}}(\overline{X}, \overline{Y}) = f(\overline{X}, \overline{Y})P_{SS} - D\partial P_{SS}/\partial \overline{X}$ and $J_{SS,\overline{Y}}(\overline{X}, \overline{Y}) = f(\overline{X}, \overline{Y})P_{SS} - D\partial P_{SS}/\partial \overline{Y}$ [33], where the subscript $SS$ represents the steady state (See details in the supplemental information, SI).

By applying above analysis on the (+) particle, the nonequilibrium landscape (colored background) and flux vectors (arrows) for typical external noise intensity $D_e$ are depicted in Figure 3a to 3f. For $D_e = 0$ (Figure 3a and 3b), two paths lying in the valleys of the nonequilibrium landscape with fluxes symmetric with respect to the X axis can be observed, corresponding to the two moving paths observed in the inset of Figure 2. However, with the increasing noise intensity, the flux becomes no more symmetric (e.g., for $D_e = 10^{-4}$ shown in Figure 3c and 3d). It is obvious that the flux vectors pointing to the positive direction are much larger than those pointing to the negative one. Consequently, the biased flux will drive (+) particles moving along the positive path, rather than the negative one, leading to complete chirality sorting ($S = 1$).

As the noise intensity further increases to be large enough (such as $D_e = 0.1$ in Figure 3e and 3f), the landscape becomes much more flat and the flux recovers to be symmetric again, implying a mixed state ($S = 0$) dominated by the noise. In short, for an appropriate intensity of the external noise, the biased-flux-induced moving direction selection of the chiral particles leads to the observed optimal chirality sorting with 100% chirality selectivity.

Herein, the origin of the biased flux occurring in the achiral potential turns into an interesting question. We find that a pair of particle with same chirality cannot move symmetrically with respect to the X axis because their bead positions, centers of friction and orientation angles are impossible to maintain symmetric simultaneously, which leads to different force or torques exerting on the particles, and eventually breaks the symmetric motion at the particle level (See details in SI). Besides, it is observed that the (+) particle always oscillates its $\phi$ below 0 periodically along the positive path (the blue line in Figure 3g), and above 0 along the negative one (the red line in Figure 3g). The ranges of oscillation along these two paths are asymmetric with respect to $\phi = 0$, indicating an intrinsic dynamical asymmetry at the trajectory level. More interestingly, a noise-induced reversible transition between the positive and negative paths can be observed for $D_e > 5.0 \times 10^{-6}$ (the black line in Figure 3g), along which the (+) particle on the negative(positive) path firstly jumps to a state with opposite sign of $\phi$ while still keeping the negative(positive) moving direction (Step ① in Figure 3g), then changes its moving direction oppositely and along the

positive(negative) one (Step ② in Figure 3g, movie S3 and S4). Thus, such a transition can be named as the multi-step transition. To present how the multi-step transition affects the particle distribution along the positive and negative paths, the probability $\rho$ of how long the (+) particle stays along the two paths is plotted in Figure 3h. When noise is absent or too large, the (+) particle moves along both of the paths with equal probabilities (green and blue bars in Figure 3h). For moderate noise intensity (such as $D_e = 10^{-4}$), the (+) particle prefers to stay much longer along the positive path than the negative one (red bars in Figure 3h), indicating a biased flux driving (+) particle moving from the negative path to the positive one at the path-transition level. Therefore, we argue that the asymmetrical flux is generated by a noise-induced path transition, which may be further associated with the intrinsic dynamical asymmetry of enantiomorphs.

We investigate whether the chirality sorting of enantiomorphs can be tuned by $D_e$ for other arrangement of values for $\gamma_i$s. Then we use the parameter $\Delta\gamma \equiv \gamma_3 - \gamma_2 = \gamma_2 - \gamma_1$ to characterize the degree of chirality for chiral particles by setting the difference between two consequent $\gamma_i$s to be the same. For $\Delta\gamma = 1.0$, i.e., $\gamma_1 = 1.0$, $\gamma_2 = 2.0$ and $\gamma_3 = 3.0$, dependence of $S$ on $D_e$ is plotted in Figure 4a. Similar to the one for $\Delta\gamma = 0.5$, $S$ is nearly 0 for $D_e < 5.0 \times 10^{-6}$. In contrast to the situation for $\Delta\gamma = 0.5$, $S$ decreases quickly to $S = -1$ as $D_e$ increases to be slightly larger than $5.0 \times 10^{-6}$, indicating a chirality sorting with all of the (+) particles only distributed below the X axis. Remarkably, by further increasing $D_e$ ($> 6.0 \times 10^{-5}$), a rollover of chirality sorting is observed, i.e., S increases rapidly from -1 to 1. In both of the parameter regions where $|S| = 1$, chiral particles are separated completely while moving path of the (+) particles is the negative one for $S = -1$ and the positive one for $S = 1$. As $D_e$ increases to be large enough, $S$ drops back to 0 again. In short, for $\Delta\gamma = 1.0$, similar noise-induced optimal chirality sorting with 100% selectivity can also be achieved, along with a new noise-induced rollover of chirality sorting.

The interesting observation can be understood based on the nonequilibrium potential landscape and flux analysis, too (Figure 4b, 4c and Figure S2). Similar to $\Delta\gamma = 0.5$, the flux is symmetric with respect to the X axis can be observed at small or large noise intensity. For noise intensity in between, it is a little complicated. As $D_e$ increases to be slightly larger than $5.0 \times 10^{-6}$ (such as $D_e = 10^{-5}$, Figure 4b), the flux to the negative direction are much larger than that to the positive one, so that the biased flux would drive (+) particles moving along the negative path, in contrast to the situation for $\Delta\gamma = 0.5$. By further increasing $D_e$ to be, such as, $D_e = 10^{-4}$ (Figure 4c), the flux to the positive direction become larger in comparison with that to the negative direction, in accordance with the rollover from $S = -1$ to $S = 1$ aforementioned. In other words, noise can further induce a change of the "flux direction" for $\Delta\gamma = 1.0$, leading to the observation of noise-induced rollover of optimal chirality sorting.

We are now interested in why the flux direction can be changed. As discussed in the chirality sorting for $\Delta\gamma = 0.5$, the biased flux originates from the intrinsic dynamical asymmetry of the chiral particles and the noise-induced transition between the moving paths. Notice that, the intrinsic dynamical asymmetry still holds for $\Delta\gamma = 1.0$, thus the transition between the moving paths is focused in following analysis. For small noise intensity (such as $D_e = 10^{-5}$), a new type of dominant reversible transition is observed. As indicated by the green line in Figure 4d, the (+) particle jumps from the positive path to the negative one directly

(named as the direct transition, movie S5 and S6). It is found that, via the direct transition, the (+) particle prefers to distribute along the negative path (red bars in Figure 4e), thus, the direct transition would result in a biased flux maining pointing to the negative direction. For large $D_e$, it is observed that a multi-step transition similar to the one for $\Delta\gamma = 0.5$ emerges and dominates the path transition process, which forces the (+) particle to distribute mainly along the positive path (blue bars in Figure 4e), leading to another type of biased flux preferring to pointing to the positive direction. In between, these is a mixed state for $D_e =$

$6.0 \times 10^{-5}$ as depicted by the nearly equal-height cyan bars in Figure 4e, where chiral particles cannot be

separated. In short, noise-shifted dominant transition between particle moving paths is the very reason for the changing of flux direction for $\Delta\gamma = 1.0$.

Based on the revealed physics, we expect to explore the possibility for noise-controlled moving

directions of chiral particles. The most probable moving directions of (+) particles

$\theta_m = \langle \lim_{t\to\infty} \text{atan}\,(Y(t)/X(t)) \rangle$ as functions of the external noise intensity $D_e$ with $\Delta\gamma = 0.5$ and 1.0 are presented in Figure 5a. It is observed that $\theta_m$ can be tuned quantitatively from 0.42 to 0 for $\Delta\gamma = 0.5$ by increasing $D_e$ (the black line in Figure 5a). Similarly, $\theta_m$ for $\Delta\gamma = 1.0$ can also be tuned from -0.42 to 0.07 then back to 0 with increasing $D_e$ (the red line in Figure 5a). In comparison with that for $\Delta\gamma = 0.5$, not only the value of $\theta_m$, but also the the sign can be tuned by solely changing $D_e$. Thus, it is possible for us to guide chiral particles to some given positions by solely tuning the intensity of external noise, which provides a more powerful tool for applications in real systems beyond the noise-induced chirality sorting.

More interestingly, one should note that different values of $|\theta_m|$ mean different final positional distributions of chiral particles, so it is possible for us to provide a potential routine to separate not only a pair of enantiomorphs with the same degree of chirality but also many kinds of enantiomorphs with different $\Delta\gamma$, simultaneously. To validate this idea, the distribution $P(Y)$ of four different chiral particles (two kinds of enantiomorphs with $\Delta\gamma = 0.5$ and $1.0$) is shown in Figure 5b and 5c for $D_e = 4.0 \times 10^{-5}$ and $10^{-4}$, corresponding to the cyan and magenta dashed lines in Figure 5a, respectively. Clearly, for both $D_e$, all of these four kinds of particles fall in different areas on the Y axis without any overlap, indicating that all these four kinds of particles can be separated perfectly at the same time.

To fully explore how parameters affect the chirality sorting, phase diagram in the $D_e$-$\Delta\gamma$ plane is obtained (Figure 6). There are several states which can be observed in the phase diagram. In the complete separation state (both of the dark blue and dark yellow domains), chiral particles can be completely separated as $|S| \approx 1$, while (+) particles move along the positive direction in the dark yellow domain and along the negative direction in the dark blue one. Chiral particles can only be partially separated in the partial separation state (the light blue and light yellow domains), and cannot be separated at all in the mixed state (the light green domain). For both $\Delta\gamma$ less than 0.8 and $\Delta\gamma$ larger than 2.5, chirality selectivity depends non-monotonically on the intensity of the external noise. Such optimal chirality sorting behavior can be observed for very large $\Delta\gamma$ (even for $\Delta\gamma > 100$). For $\Delta\gamma$ in between, noise-induced rollover of chirality selectivity occurs. This complete phase diagram consisting of various states enables feasibilities for many new chirality sorting routines in practice.

To explore the generalizability of the noise-tuned chirality sorting, chirality sorting in different types of periodic potentials and with different types of noises are also investigated by intensive simulations. Figure 7a and 7b show the chirality selectivity $S$ as functions of noise intensity $D_e$ in two periodic potentials of other shapes (see details in SI). Obviously, an optimal chirality sorting with 100% selectivity can also be induced by the external noise in these potentials. We have also observed that chirality sorting won't happen when the potential is uniform along the direction parallel or vertical to the flowing fluid. Combined with the potential lying on interpenetrating triangular lattices aforementioned, we argue that chirality sorting could still be observed if the potential is periodic both along the X axis (the flowing direction) and along the Y axis (vertical to the flowing direction) while other details of the potential such as the shape seems to be irrelevant.

To further validate the concept of noise-tuned chirality sorting in systems of correlated noise (experimentally controlled noise is usually correlated), similar simulations are performed by replacing the external white noise by time-correlated or space-correlated noise (see SI for details). The external time-correlated noise $\boldsymbol{g}(t)$ is chosen as Ornstein-Uhlenbeck noise whose correlation function $\langle \boldsymbol{g}(t)\boldsymbol{g}(t + \Delta t)\rangle = D_e e^{-\Delta t/\tau_p}$ (inset of Figure 7c), where $\tau_p$ is the characteristic time scale of the correlation. The space-correlated noise requires the construction of a random noise field $\Phi(\boldsymbol{x}, t)$ (inset of Figure 7d)[34]. As expected, chirality sorting and its rollover can be observed with these kinds of noises, too (Figure 7c and 7d). Besides, the finite correlation time(scale) of such correlated noise further offers a rich parameter space for tunable chirality sorting, which may be systematically investigated in future works.

**Conclusion**

In summary, we found an optimal sorting of mesoscopic chiral particles occurring with the help of external noise in an achiral periodic potential. The moving direction of chiral particles can be well-controlled by the intensity of external noise, leading to a conceptually new noise-tuned chirality-sorting method. Such interesting effect originates from a biased flux driving particles to move along a selected direction, which is generated by the noise-induced path transition. The robustness and generalizability of noise-tuned chirality sorting were further demonstrated in systems with other types of potentials or spatially/temporally correlated noise. Compared with conventional separation methods, the noise-tuned sorting of mesoscopic chiral structures, on the one hand, broadens our ability of chirality sorting for practical purposes where other methods may fail, and on the other hand, provides a new basic concept for tunable chirality sorting by utilizing the constructive role of external noise. For instance, we demonstrated that, based on the noise-tuned moving directions of chiral particles, it is possible to simultaneously separate several types of enantiomorphs just by adapting the intensity of external noise solely. Since external noise is independent on the internal properties of the systems and can be conveniently controlled than internal noise, our method may open a brand-new perspective on both theoretical and experimental investigations of tunable sorting of chiral structures assembled from micro- or meso-blocks in the future.

**Supplementary Materials**

Figure S1. The intrinsic dynamical asymmetry of enantiomorphs.

Figure S2. Nonequilibrium landscape and flux vectors of the (+) particle.

Figure S3. Correlated external noise induced chirality sorting for particles with different chirality degrees.

Figure S4. A typical snapshot of the 2D space-correlated noise field for chiral particles.

Movie S1. The positive path for the (+) particle.

Movie S2. The negative path for the (+) particle.

Movie S3. First type of the multi-step transition for the (+) particle.

Movie S4. Second type of the multi-step transition for the (+) particle.

Movie S5. First type of the direct transition for the (+) particle.

Movie S6. Second type of the direct transition for the (+) particle.

**Corresponding Author**

*E-mail: hzhlj@ustc.edu.cn

*E-mail: hjjiang3@ustc.edu.cn

**ORCID**

Jie Su: 0000-0002-5413-5790

Hui-Jun Jiang: 0000-0001-7243-5431

Zhong-Huai Hou: 0000-0003-1241-7041

**Author Contributions**

H.-J.J. and Z.-H.H. conceived the idea and supervised the research, J.S. performed the simulations and data analysis. All authors discussed the results and underlying mechanism, and co-wrote the manuscript.

**Notes**

The authors declare no competing financial interest.

**Acknowledgement**

We acknowledge valuable discussions with Prof. Jin Wang. This work is supported by MOST (2016YFA0400904, 2018YFA0208702), by NSFC (21833007, 21790350, 21673212, 21521001, 21473165), by the Fundamental Research Funds for the Central Universities (2340000074), and Anhui Initiative in Quantum Information Technologies (AHY090200).

**Figures**

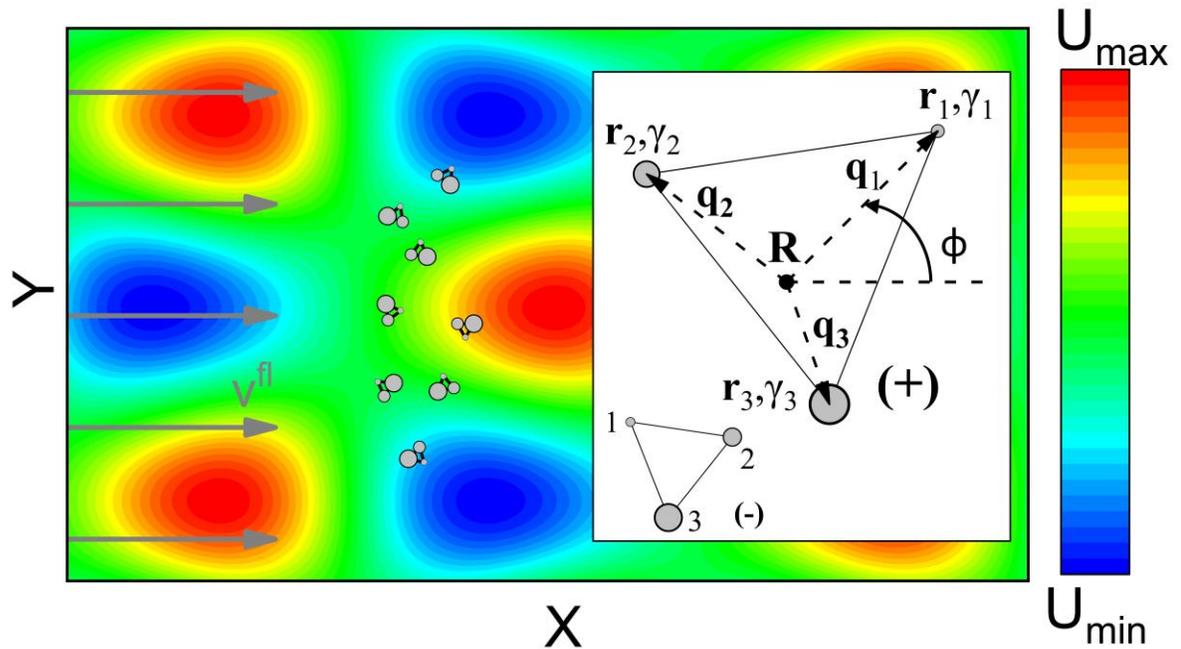

**Figure 1.** Schematic of the chiral particles driven by a fluid field $v^{fl}$ (gray arrows) through an achiral periodic potential $U(r)$ (colored background). The inset is the zoom-in of a pair of chiral particles. The (+) particle consists of three nanoparticles arranged counterclockwise with $\gamma_1 < \gamma_2 < \gamma_3$ on an equilateral triangle's vertexes $\boldsymbol{r_i}$. $\boldsymbol{R}$ is the center of friction and $\phi$ is the angle between the vector $\boldsymbol{r_1} - \boldsymbol{R}$ and the X axis. The (-) particle differs only in that the sequence of $\gamma_i$s is clockwise.

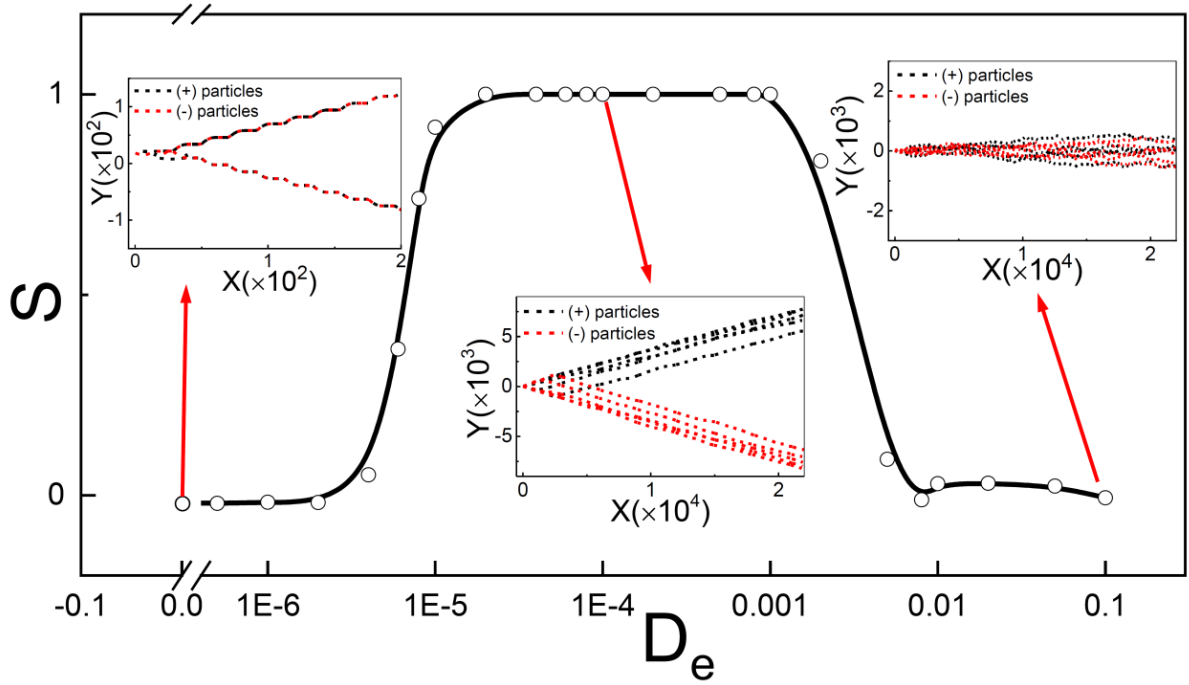

**Figure 2.** Optimal chirality sorting induced by the external noise. Chirality selectivity $S$ as a function of the external noise intensity $D_e$ for $\gamma_1 = 1.0$, $\gamma_2 = 1.5$ and $\gamma_3 = 2.0$. The top-left, bottom and top-right insets are typical trajectories of the ($\pm$) particles for $D_e = 0$, $10^{-4}$ and $0.1$, respectively.

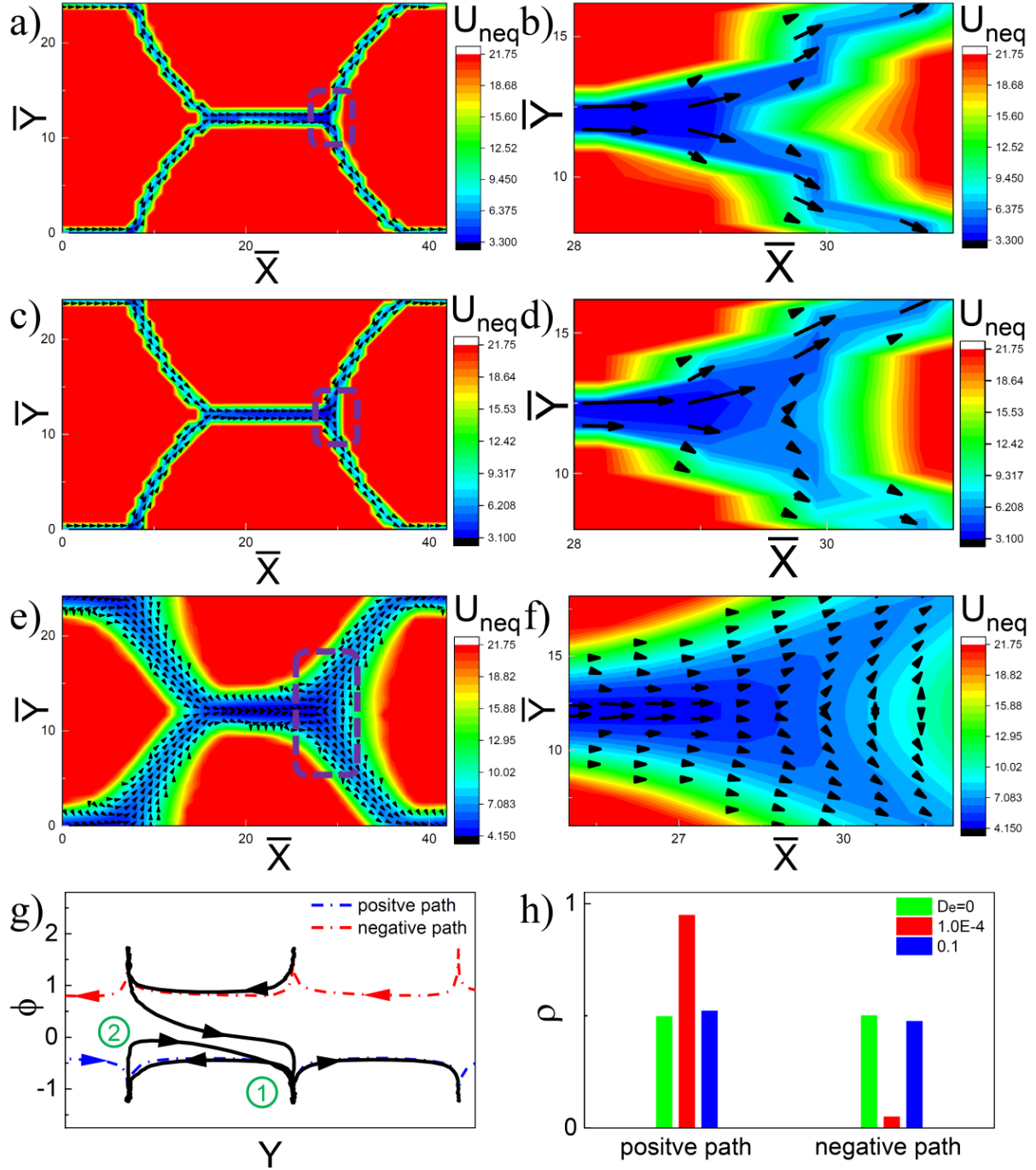

**Figure 3.** Underlying mechanism for the noise-induced chirality sorting. (a), (c), (e) Nonequilibrium landscape (colored background) and flux vectors (arrows representing the directions of flux) of the (+) particle for external noise intensity $D_e = 0$, $10^{-4}$, $0.1$ (b), (d), (f) Zoom-in of the purple boxes in (a), (c), (e), respectively (the length of arrows representing the strength of flux). (g) Dominant transition (the multi-step transition) of (+) particles from the negative to the positive path. The blue and red lines show the positive and the negative path, respectively. Arrows are the moving direction of particles. (h) The probabilities $\rho$ of how long the (+) particles stay in the positive and negative paths.

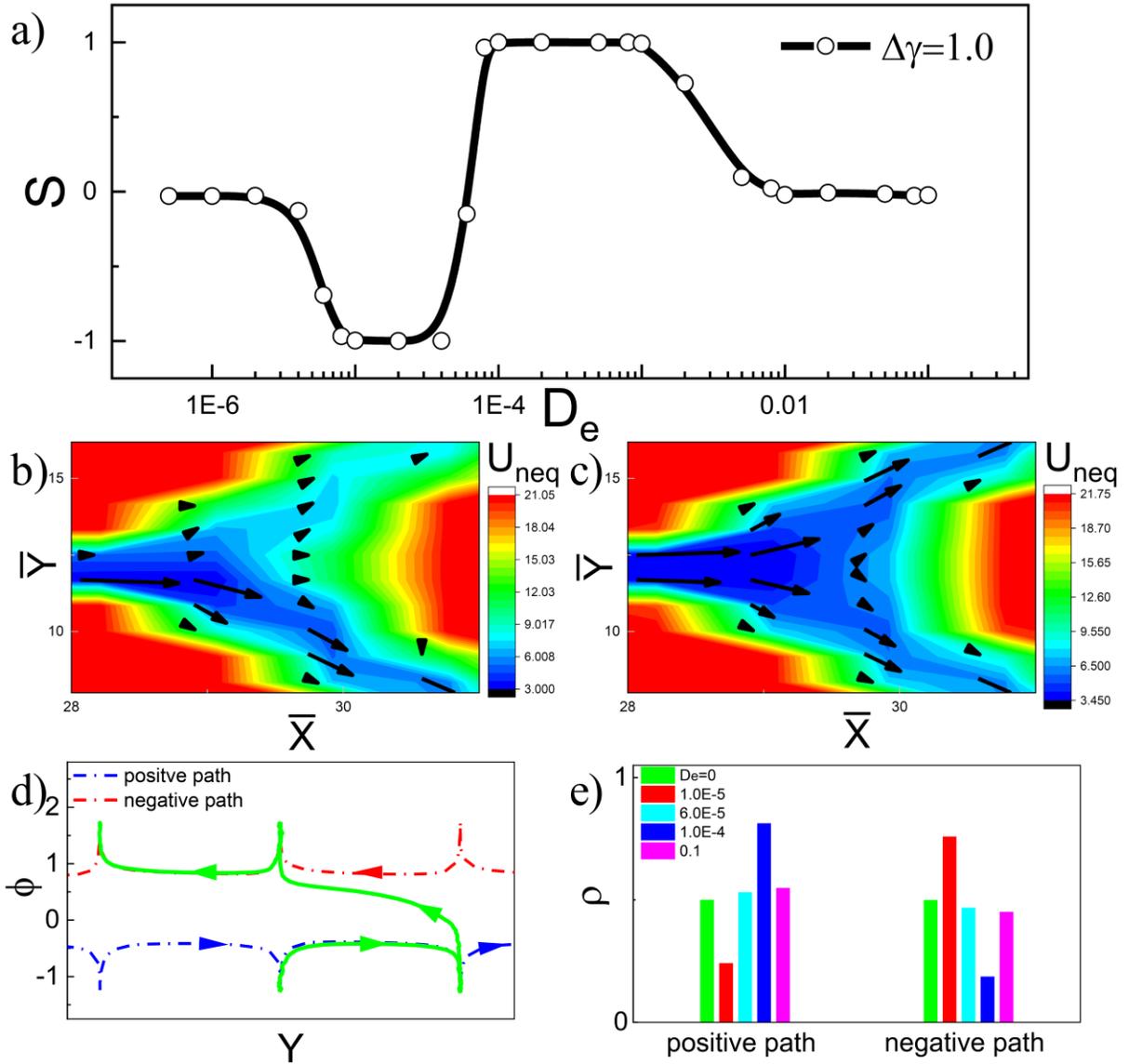

**Figure 4.** Chirality sorting and underlying mechanism for chirality sorting with $\Delta\gamma = 1.0$. (a) Chirality selectivity $S$ as a function of the external noise intensity $D_e$. (b), (c) The nonequilibrium landscape (colored background) and flux vectors (arrows and its length representing the directions and strength of flux) of the (+) particle with $\Delta\gamma = 1.0$ for external noise intensity $D_e = 10^{-5}, 10^{-4}$. (d) Dominant transition (the direct transition) of (+) particles from the negative to the positive path for $S = -1$. The blue and red lines show the positive and the negative path, respectively. Arrows are the moving direction of particles. (e) The probabilities $\rho$ of how long the (+) particles stay in the positive and negative paths for $\Delta\gamma = 1.0$.

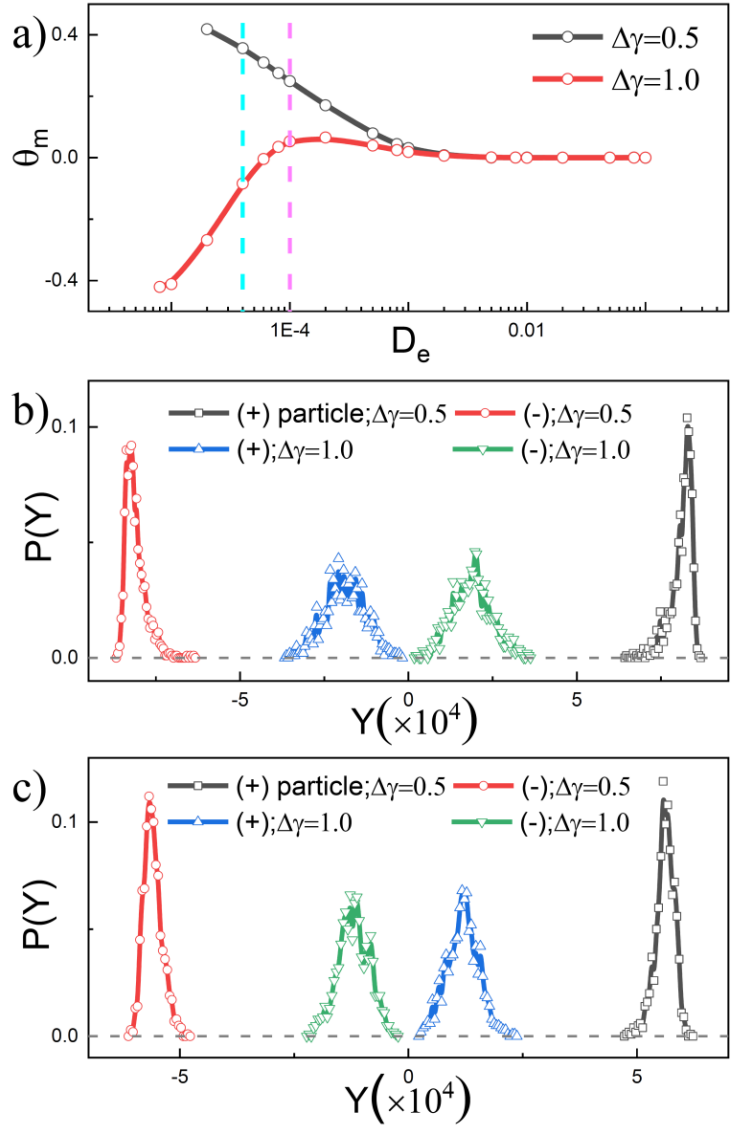

**Figure 5.** Noise-induced tunable chirality sorting. (a) The most probable moving direction $\theta_m$ of the (+) particle for $\Delta\gamma = 0.5$ and 1.0. (b), (c) Simultaneous chirality sorting for different kinds of enantiomorphs. Distributions on the Y-position for four different chiral particles with $\Delta\gamma = 0.5$ and 1.0 for external noise intensity $D_e = 4.0 \times 10^{-5}$ and $1.0 \times 10^{-4}$ (cyan and magenta dashed lines in (a)).

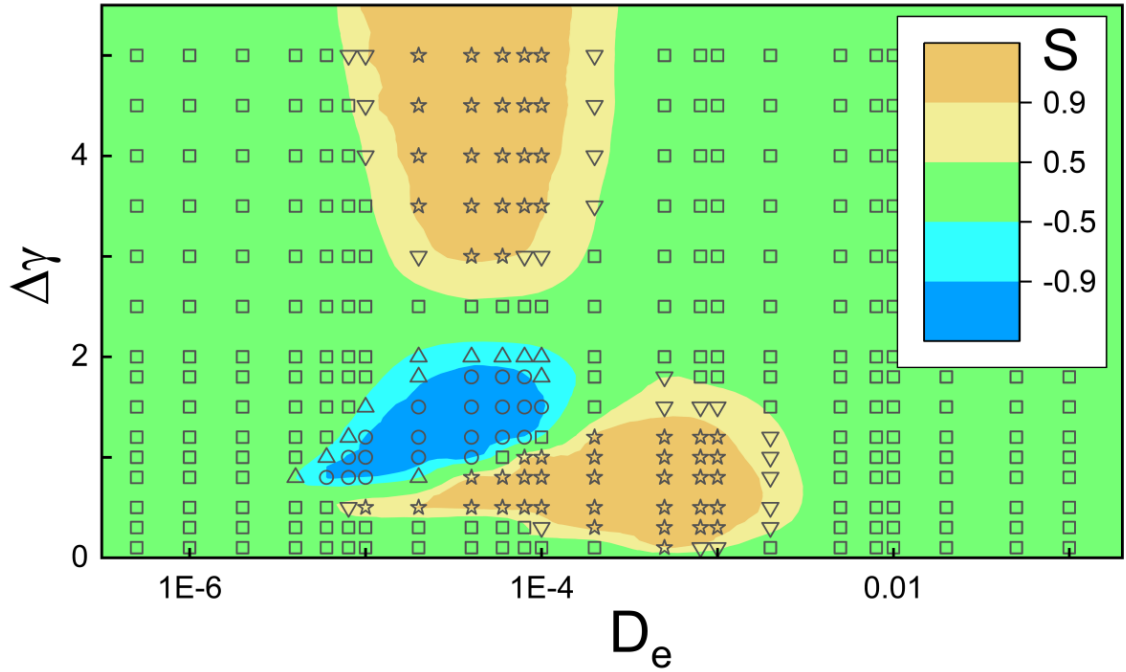

**Figure 6.** The phase diagram for chirality sorting in the $D_e$-$\Delta\gamma$ plane. Circles and stars: complete separation state. Triangles: partial separation state. Squares: mixed state. The (+) particles prefer to move along the positive direction in yellow domains, while tend to the negative one in blue domains.

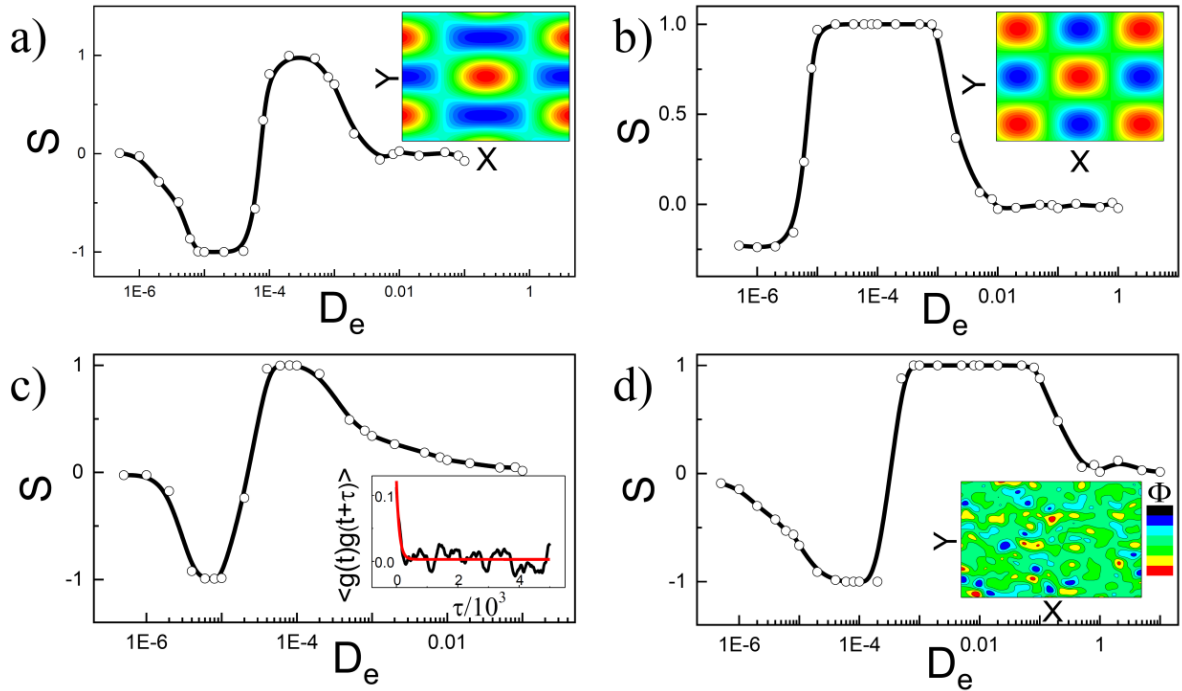

**Figure 7.** Generalizability of noise-tuned chirality sorting. Chirality selectivity $S$ as functions of $D_e$ with $\Delta\gamma = 1.0$ for (a), (b) periodic potentials with different shapes, and (c) time- or (d) space-correlated noises. Insets in (a), (b): periodic potentials of different shapes. Inset in (c): correlation function of the time-correlated noise (the black line) and its exponential fit (the red line). Inset in (d): a typical snapshot of 2D random space-correlated noise field. Corresponding value of the fields in insets of (a), (b) and (d) increases as the color changes from blue to red.

**TOC Grapgic**

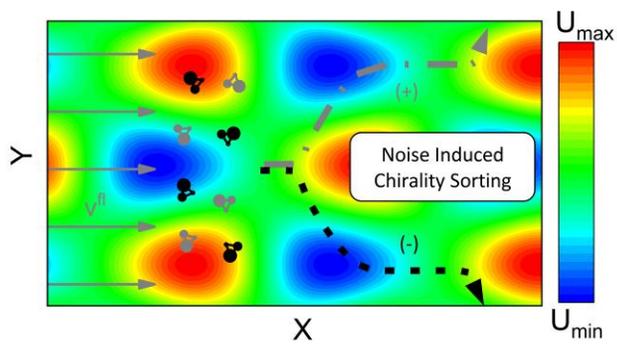

Noise Induced Chirality Sorting